\documentclass[%
preprint,superscriptaddress,
amsmath,amssymb,prd,aps,showpacs
floatfix,
]{revtex4-1}
\usepackage{graphicx}
\usepackage{dcolumn}
\usepackage{bm}
\usepackage{mathrsfs}
\usepackage{amsmath}
\usepackage{amssymb}
\usepackage{float}
\usepackage{dcolumn}
\usepackage{multirow}

\begin{document}


\title{Residues of $\Lambda_Q$-type and $\Sigma_Q$-type Baryons in the Bethe-Salpeter Equation Approach}

\author{Qi-Xin Yu}
\email[E-mail: ]{yuqx@mail.bnu.edu.cn}
\affiliation{College of Nuclear Science and Technology,
Beijing Normal University, Beijing 100875, China}

\author{Zhen-Yang Wang}
\affiliation{College of Nuclear Science and Technology,
Beijing Normal University, Beijing 100875, China}

\author{Jing-Juan Qi}
\affiliation{College of Nuclear Science and Technology,
Beijing Normal University, Beijing 100875, China}

\author{Xin-Heng Guo}
\email[Corresponding author, e-mail: ]{xhguo@bnu.edu.cn}
\affiliation{College of Nuclear Science and Technology,
Beijing Normal University, Beijing 100875, China}


\begin{abstract}
We study the residues of $\Lambda_Q$-type baryons ($\Lambda_Q$ and $\Xi_Q^A$) $(Q=b,c)$ and $\Sigma_Q$-type baryons ($\Sigma_Q^{(\ast)}$, $\Xi_Q^{S(\ast)}$ and $\Omega_Q^{(\ast)}$) in the quark-diquark model within the Bethe-Salpeter (BS) formalism. These residues can be used, for example, in the calculations of the amplitudes in the scattering processes. After constructing the baryonic currents in the BS formalism, we derive the relations between the BS wave functions and the residues for these baryons. The BS equations are solved numerically with the kernel including the scalar confinement and the one gluon exchange terms and with the covariant instantaneous approximation being employed in the calculations. Finally, we obtain the numerical values of the residues $0.103\,\rm GeV\sim0.224\,\rm GeV$ for $\Lambda_Q$, $0.143\,\rm GeV\sim0.215\,\rm GeV$ for $\Xi_Q^A$, $0.262\,\rm GeV\sim0.361\,\rm GeV$ for $\Sigma_Q^{(\ast)}$, $0.313\,\rm GeV\sim0.460\,\rm GeV$ for $\Xi_Q^{S(\ast)}$ and $0.473\,\rm GeV\sim0.571\,\rm GeV$ for $\Omega_Q^{(\ast)}$ in the ranges of the parameters in our model.
\end{abstract}

\maketitle


\section{Introduction}
The residues of baryons are important parameters in studying their hadronic properties, for example, in QCD sum rules, the residues are needed as one of the main inputs for further calculating decay constants and coupling constants of baryons \cite{Wang:2015kua,Ebert:1995fp,Groote:1996em}. In this paper, we will present the calculations of the residues for $\Lambda_Q$-type $(Q=b,c)$ baryons which include $\Lambda_Q$ and $\Xi_Q^A$, and $\Sigma_Q$-type baryons which are $\Sigma_Q^{(\ast)}$, $\Xi_Q^{S(\ast)}$ and $\Omega_Q^{(\ast)}$ \cite{Guo:1998ef,Guo:1996jj}. In the heavy quark effective theory (HQET), baryons are categorized based on the angular momentum and parity $(j^P)$ of the light part. For the ground states of heavy baryons, $j^P$ of the light part can be either $0^+$ or $1^+$, which naturally gives the $\Lambda_Q$-type and $\Sigma_Q$-type baryons, respectively. Furthermore, after adding the heavy quark in these baryons, $j^P=\frac{1}{2}^+$ for $\Lambda_Q$ and $\Xi_Q^A$, and the degenerate states $j^P=\frac{1}{2}^+$, $j^P=\frac{3}{2}^+$ for $\Sigma_Q$ ($\Xi_Q^S$, $\Omega_Q$), $\Sigma_Q^\ast$ ($\Xi_Q^{S\ast}$, $\Omega_Q^\ast$) can be obtained.

The properties of baryons which contain a single heavy quark can be simplified due to fact that the mass (or the flavor) and spin of the heavy quark become irrelevant to the leading order in $1/m_Q$ expansion. Since the light part has good spin and isospin numbers,  these baryons can be investigated in the quark-diquark picture \cite{Artru:1989zv,Ebert:1995fp,Guo:1992tt,Guo:1998ef,Guo:2007qu,Meyer:1990fr,Tong:1999qs}. The residues of baryons were mainly investigated with the method of QCD sum rules during the past few decades \cite{Shuryak:1981fza,Grozin:1992td,Yakovlev:2000uc,Groote:1996em,Groote:1997ci,Wang:2015kua}. The constructed baryonic currents employed in QCD sum rules are not based on the quark-diquark model, which can be seen from the dimension of the residues \cite{Grozin:1992td}. In this work, we will construct the baryonic currents for $\Lambda_Q$-type and $\Sigma_Q$-type baryons respectively. Then the Bethe-Salpeter (BS) equation will be employed to calculate the residues of $\Lambda_Q$-type and $\Sigma_Q$-type baryons in the quark-diquark picture in the heavy quark limit $m_Q\to\infty$. These residues reflect the inner structures of heavy baryons in the quark-diquark picture.

\begin{figure}[h!]
\scalebox{0.7}[0.7]{\includegraphics{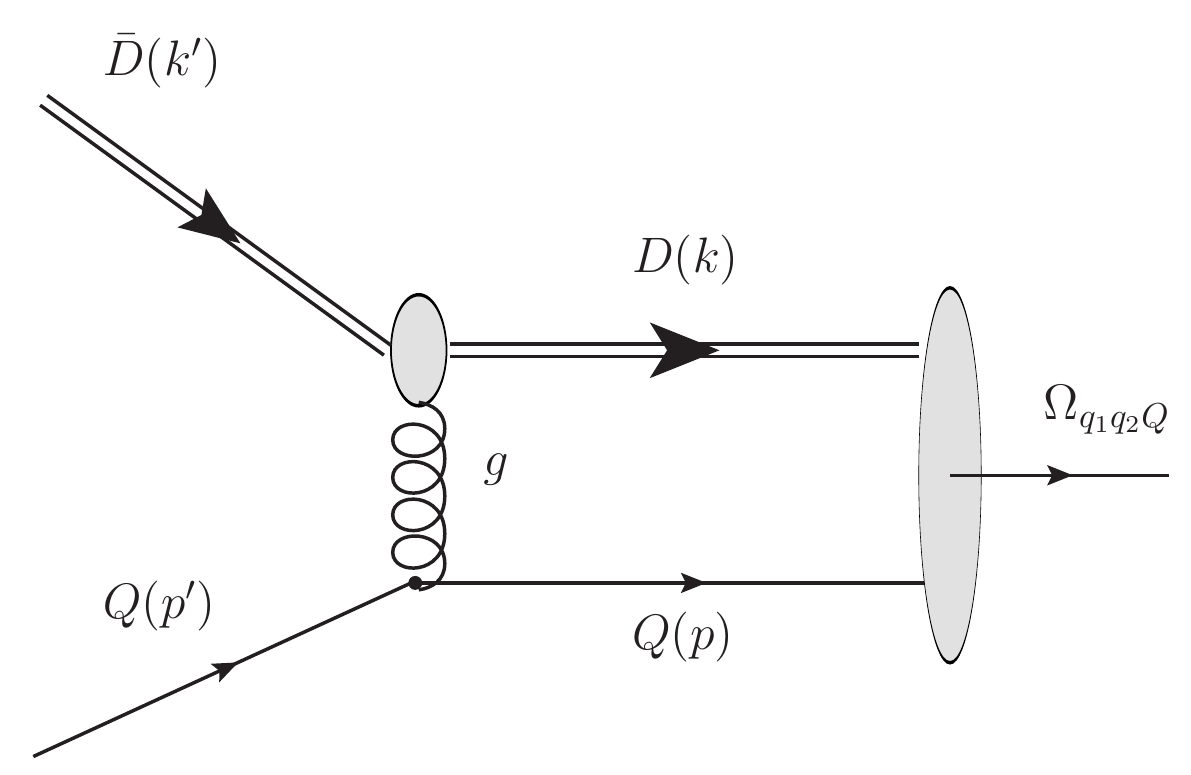}}
\caption{The example of the production of $\Omega_{q_1q_2Q}$ from a scalar diquark $D(q_1q_2)$ and a heavy quark $Q$ through the exchange of the gluon $g$ \cite{GomshiNobary:2007ofo}.}\label{fig}
\centering
\end{figure}

In order to obtain further prospects of $\Lambda_Q$-type and $\Sigma_Q$-type baryons for experimental study and future colliders, it is important to have accurate predictions for the production rates of heavy baryons, and that is where we will need the residues of these baryons. In Refs.~\cite{GomshiNobary:2009zz,GomshiNobary:2007ofo}, the authors established the connection between the baryonic current in the quark-diquark model and the scattering amplitude of a diquark and a heavy quark to a heavy baryon $\Omega_{Q^\prime Q^{\prime\prime}Q}$, where baryons containing a heavy scalar diquark and a heavy quark were considered. In our work, we will study the baryons containing one heavy quark and a light diquark. Following the formalism in Ref.~\cite{GomshiNobary:2007ofo}, the production amplitude of $\Omega_{q_1q_2Q}$ from a scalar diquark $D$ and a heavy quark $Q$ is (see Fig.~\ref{fig}, $q_1$ and $q_2$ are light quarks which compose of the diquark $D$)
\begin{equation}\label{js}
  T\sim g_s^2\frac{1}{q^2}F_S(q^2)\langle \Omega_{q_1q_2Q}|J|0\rangle,
\end{equation}
where $J$ is the baryonic current, $q$ is the momentum transfer $(q=k-k^\prime)$, $g_s$ is the strong coupling constant, and $F_S$ is the form factor of the diquark which is introduced to describe the interaction between the diquark and the gluon \cite{GomshiNobary:2007ofo}. It can be seen from Eq.~\eqref{js} that the residue of the baryon, which is related to the matrix element $\langle\Omega_{q_1q_2Q}|J|0\rangle$, is needed in the calculation. Conversely, although there has been no direct experimental measurement of the residues for $\Lambda_Q$-type and $\Sigma_Q$-type baryons, one can make use of the relation between the residue in the quark-diquark model and the scattering amplitude for the production of the heavy baryon in Eq.~\eqref{js} to obtain the values of the residues of heavy baryons from experimental data. Obviously, the residues of heavy baryons can also be used to calculate fragmentation functions of heavy baryons.

The BS equations for the $\Lambda_Q$-type and the $\Sigma_Q$-type baryons have been established in the quark-diquark picture in the leading order of the $1/m_Q$ expansion \cite{Guo:1996jj,Guo:1998ef}. They were applied to study the properties of the $\Lambda_Q$-type and the $\Sigma_Q$-type baryons, and theoretical predictions were found to be consistent with the available experimental data \cite{Guo:2007qu,Liu:2018tqe}. The kernel for the BS equation is motivated by the potential model, which consists of the scalar confinement and the one gluon exchange terms \cite{Eichten:1978tg,Guo:1998ef,Guo:2007qu,Liu:2018tqe,Wei:2015gsa,Salpeter:1951sz}. We will construct the formalism for the residue calculations in the BS equation approach \cite{Lurie:1968zz}. After that, the BS equations will be solved using the covariant instantaneous approximation \cite{Eichten:1978tg,Jin:1992mw,Dai:1993np,Dai:1993kt}, and the obtained BS wave functions will be applied to calculate the residues of $\Lambda_Q$, $\Xi_Q^A$, $\Sigma_Q^{(\ast)}$, $\Xi_Q^{S(\ast)}$ and $\Omega_Q^{(\ast)}$.

This paper is organized as follows. In Sec.~\uppercase\expandafter{\romannumeral2}, we will construct the baryonic currents for $\Lambda_Q$-type and $\Sigma_Q$-type baryons, and define the residues of the heavy baryons in the quark-diquark picture. In Sec.~\uppercase\expandafter{\romannumeral3}, the BS equations for $\Lambda_Q$-type and $\Sigma_Q$-type baryons will be presented under the heavy quark limit. In Sec.~\uppercase\expandafter{\romannumeral4}, we will give the numerical results for the BS wave functions of $\Lambda_Q$-type and $\Sigma_Q$-type baryons and their corresponding residues. Finally, the summary will be given in Sec.~\uppercase\expandafter{\romannumeral5}.
\section{Baryonic currents for $\Lambda_Q$-type and $\Sigma_Q$-type baryons }
In this section, we will present the baryonic currents for the heavy baryons. Generally, for a baryon composed of a heavy quark $Q$ and two light quarks $q_1$ and $q_2$, the baryonic current has the form \cite{Grozin:1992td,Groote:1996em,Groote:1997ci}:
\begin{equation}\label{general form}
  J=[q_1^{iT}C\Gamma\tau q_2^j]\Gamma^\prime Q^k\varepsilon_{ijk},
\end{equation}
where $i,j,k$ are the color indices, the index $T$ indicates the transposition of the matrix, $C$ stands for the charge conjugation matrix, $\tau$ is the matrix in the flavor space and $\Gamma$ and $\Gamma^\prime$ represent the light side and heavy side Dirac structures of current vertices, respectively. Neglecting color indices and $\tau$, the current could be abbreviated as the form
\begin{equation}\label{abbrev from}
  J=[q^TC\Gamma q]\Gamma^\prime Q.
\end{equation}
Then, the current for the $\Lambda_Q$-type baryon including the scalar diquark $D$ $(q^TC\gamma^5q)$ composed of two light quarks can be written in the following form:
\begin{equation}\label{sc}
  J_{\Lambda_Q}=DQ.
\end{equation}
In the rest frame, $\Sigma_Q$-type and $\Sigma_Q^{\ast}$-type baryons are described by the current $\vec J=\vec AQ$ with $\vec A$ being the isovector diquark field \cite{Grozin:1992td,Georgi:1990cx,Falk:1991nq}. $\vec J$ can be decomposed as
\begin{equation}\label{sigma}
  \vec J=\vec AQ=\left(\vec J+\frac{1}{3}\vec\gamma\vec\gamma\cdot\vec J\right)-\frac{1}{3}\vec\gamma\vec\gamma\cdot\vec J,
\end{equation}
where the part $\vec J+\frac{1}{3}\vec\gamma\vec\gamma\cdot\vec J$ satisfies the relation $\vec\gamma\cdot\vec J=0$, thus has spin $3/2$ corresponding to the $\Sigma_Q^{\ast}$-type baryon, and the other part can be written as the following form:
\begin{equation}
  -\frac{1}{3}\vec\gamma\vec\gamma\cdot\vec J=-\frac{1}{3}\vec\gamma\gamma_5\gamma_5\vec\gamma\cdot\vec AQ=\frac{1}{3}\vec\gamma\gamma_5J_{1/2},
\end{equation}
where $J_{1/2}=\vec A\cdot\vec\gamma\gamma_5Q$ satisfies the condition $\gamma_0\vec J_{1/2}=\vec J_{1/2}$, indicating $\vec J_{1/2}$ has spin $1/2$ and corresponds to $\Sigma_Q$. The above currents can be generalized to the general frame by substituting $\gamma_0$ with $v\!\!\!/$ and $\vec\gamma$ with $\gamma^\bot_\mu=\gamma_\mu-v\!\!\!/v^\mu$ ($v$ is the velocity of $\Sigma_Q$-type or $\Sigma_Q^{\ast}$-type baryon). With these substitutions, we arrive at the current:
 \begin{equation}
   J_{1/2}=A_\mu(\gamma^\mu-v\!\!\!/v^\mu)\gamma_5Q.
\end{equation}
Since $A_\mu$ satisfies the condition $v^\mu A_\mu=0$, we have the following form for the baryonic current of $\Sigma_Q$-type baryon
\begin{equation}
  J_{\Sigma_Q}=J_{1/2}=A_\mu\gamma^\mu\gamma_5Q.
\end{equation}
Similarly, we have the following form for the current of $\Sigma_Q^\ast$-type baryon:
\begin{equation}
  J^\mu_{\Sigma_Q^{\ast}}=A^\mu Q+\frac{1}{3}(\gamma^\mu-v\!\!\!/v^\mu)A\!\!\!/Q.
\end{equation}
We can define the residues for $\Lambda_Q$-type and $\Sigma_Q^{(\ast)}$-type baryons in the quark-diquark picture in the following forms:
\begin{equation}\label{def1}
  \langle 0|J_{\Lambda_Q}|\Lambda_Q\rangle = f_{\Lambda_Q}u_{\Lambda_Q},
\end{equation}
\begin{equation}\label{def2}
  \langle 0|J_{\Sigma_Q}|\Sigma_Q\rangle = f_{\Sigma_Q}u_{\Sigma_Q},
\end{equation}
\begin{equation}\label{def3}
  \langle 0|J^\mu_{\Sigma^\ast_Q}|\Sigma^\ast_Q\rangle = \frac{1}{\sqrt 3}f_{\Sigma^\ast_Q}u^\mu_{\Sigma^\ast_Q},
\end{equation}
where $f_{\Lambda_Q}$, $f_{\Sigma_Q}$ and $f_{\Sigma_Q^\ast}$ are residues for $\Lambda_Q$-type, $\Sigma_Q$-type and $\Sigma_Q^\ast$-type baryons, respectively, $u_{\Lambda_Q}$, $u_{\Sigma_Q}$ and $u^\mu_{\Sigma_Q^\ast}$ are their corresponding spinors. We will show later that in the leading order of HQET the residues for the spin doublet, for example, $\{\Sigma_Q, \Sigma_Q^\ast\}$, are the same.
\section{The Bethe-Salpeter formalism for $\Lambda_Q$-type and $\Sigma_Q$-type baryons}
\subsection{The BS equations for $\Lambda_Q$-type baryons}
In this subsection, we will take $\Lambda_Q$ as an example for the $\Lambda_Q$-type baryons. $\Lambda_Q$ can be treated as the composition of a light scalar diquark and a heavy quark. As mentioned before, in this quark-diquark picture, the BS wave function for $\Lambda_Q$ is defined as \cite{Guo:1996jj}
\begin{equation}\label{BSlambda}
  \chi(x_1,x_2,P)=\langle 0|T\psi(x_1)\phi(x_2)|\Lambda_Q(P)\rangle,
\end{equation}
where $P=m_{\Lambda_Q}v$ is the momentum of $\Lambda_Q$, and $v$ is its velocity, $\phi(x_1)$ and $\psi(x_2)$ are the field operators of the scalar diquark and the heavy quark, respectively. Define $\lambda_1=\frac{m_Q}{m_Q+m_D}$, $\lambda_2=\frac{m_D}{m_Q+m_D}$, where $m_Q$ and $m_D$ are the masses of the heavy quark and the light diquark, respectively. The BS equation in momentum space can be expressed as
\begin{equation}\label{equation}
  \chi(x_1,x_2,P)=e^{-iPX}\int\frac{d^4p}{(2\pi)^4}e^{-ipx}\chi_P(p),
\end{equation}
where $X=\lambda_1x_1+\lambda_2x_2$ is the coordinator of center of mass, $x=x_1-x_2$ is the relative coordinator, and $p$ is the corresponding relative momentum. Then the momentum of the heavy quark is $p_1=\lambda_1P+p$, and $p_2=\lambda_2P-p$ for the light diquark.

The BS equation for $\Lambda_Q$ has the following form \cite{Guo:1996jj}:
\begin{equation}\label{bse}
  \chi_P(p)=S_F(p_1)\int\frac{d^4q}{(2\pi)^4}G(P,p,q)\chi_P(q)S_D(p_2),
\end{equation}
where $G(P,p,q)$ is the kernel which includes the one gluon exchange term and the confinement term \cite{Eichten:1978tg,Guo:1998ef},
\begin{equation}\label{kernel}
  -iG=I\otimes IV_1+v_\mu\otimes(p_2+p_2^\prime)^\mu F(Q^2)V_2,
\end{equation}
where the first term corresponds to the scalar confinement while the second term is from the one gluon exchange diagram, and the vertex describing the interaction between the gluon and two scalar diquarks is $(p_2+p_2^\prime)^\mu F(Q^2)$ with $F(Q^2)$ the form factor due to the structure of the scalar diquark.

It has been shown that in the leading order of $1/m_Q$ expansion we only need one scalar function, $f_P(p)$, to describe the BS wave function of $\Lambda_Q$ \cite{Guo:1996jj}. $f_P(p)$ has the following relation with $\chi_P(p)$:
\begin{equation}\label{wave}
  \chi_P(p)=f_P(p)u_{\Lambda_Q},
\end{equation}
where $u_{\Lambda_Q}$ is the Dirac spinor of $\Lambda_Q$ and $f_P(p)$ is the scalar function of $p_l$ and $p_t^2$, which are defined as $p_l=v\cdot p-\lambda_2m_{\Lambda_Q}$, $p_t=p-(v\cdot p)v$. Substituting the explicit forms of propagators of the scalar diquark and the heavy quark into Eq.~\eqref{bse}, and completing the integral $\int\frac{dq_l}{2\pi}$, we obtain the following equation for the scalar function $\widetilde f_P(p_t)(=\int\frac{dq_l}{2\pi}f_P(p))$:
\begin{equation}\label{bsforlambda}
  \widetilde f_P(p_t)=-\frac{1}{2W_p(\lambda_1m_{\Lambda_Q}-\omega_Q-W_p)}\int\frac{d^3q_t}{(2\pi)^3}(\widetilde V_1-2W_p\widetilde V_2)\widetilde f_P(q_t),
\end{equation}
where $\omega_Q=\sqrt{m_Q^2-p_t^2}$, $W_p=\sqrt{m_D^2-p_t^2}$ and $\widetilde V_{1(2)}=\int\frac{dq_l}{2\pi}V_{1(2)}$.
The covariant instantaneous approximation is applied in obtaining Eq.~\eqref{bsforlambda}. Meanwhile, we have the explicit forms for $\widetilde V_1$,
\begin{equation}\label{V1}
  \widetilde V_1=\frac{8\pi\kappa}{[(p_t-q_t)^2+\mu^2]^2}-(2\pi)^3\delta^3(p_t-q_t)\int\frac{d^3k}{(2\pi)^3}\frac{8\pi\kappa}{(k^2+\mu^2)^2},
\end{equation}
and $\widetilde V_2$ combining the form factor $F(Q^2)=\frac{\alpha_{seff}Q_0^2}{Q^2+Q_0^2}$ \cite{Anselmino:1987vk},
\begin{equation}\label{V22}
  \widetilde V_2=-\frac{16\pi}{3}\frac{\alpha^2_{seff}Q_0^2}{[(p_t-q_t)^2+\mu^2][(p_t-q_t)^2+Q_0^2]},
\end{equation}
where $\kappa$ and $\alpha_{seff}$ are coupling parameters related to the confinement and one gluon exchange interactions, and $\mu$ is introduced to avoid the infrared divergence in numerical calculations, we will apply the limit $\mu\to 0$ in the end. $Q_0^2$ is a parameter which is used to freeze $F(Q^2)$ when $Q^2$ is small. From the analysis of the electromagnetic form factor for proton, $Q_0^2$ was found to be consistent with the experiment data when $Q_0^2=3.2\,\rm GeV^2$ \cite{Anselmino:1987vk,Liu:2018tqe}.

From the fact that the Isgur-Wise function is normalized to one at the zero recoil point, we can determine the normalization constant for $\Lambda_Q$ \cite{Guo:1996jj}. In the heavy quark limit, the Isgur-Wise function $\xi(\omega)$ ($\omega=v\cdot v^\prime$ is the velocity transfer) is related to the matrix $\langle\Lambda_c(v^\prime)|\bar c\gamma_\mu b|\Lambda_b(v)\rangle$, which has the following relation with the BS wave function of $\Lambda_Q$ \cite{Guo:1996jj}:
\begin{equation}
  \langle\Lambda_c(v^\prime)|\bar c\gamma_\mu b|\Lambda_b(v)\rangle=\int\frac{d^4p}{(2\pi)^4}\bar\chi_{P^\prime}(p^\prime)\gamma_\mu\chi_P(p)S^{-1}_D(p_2),
\end{equation}
where $v$ and $v^\prime$ are the velocities of $\Lambda_b$ and $\Lambda_c$, respectively. Then, we have
\begin{equation}\label{xi}
  \xi(\omega)=\int\frac{d^4p}{(2\pi)^4}f_{P^\prime}(p^\prime)f_{P}(p)S^{-1}_D(p_2).
\end{equation}
\subsection{The BS equations for $\Sigma_Q$-type baryons}
In this subsection, we will take $\Sigma_Q^{(\ast)}$ as an example for $\Sigma_Q$-type baryons. $\Sigma_Q^{(\ast)}$ is regarded as a bound state of an axial-vector diquark $A_\mu$ and a heavy quark $\psi_Q$. We can use $B_\mu=u_Q\epsilon_\mu$ to represent the heavy baryon state, where $u_Q$ stands for the Dirac spinor of $Q$, and $\epsilon_\mu$ is the polarization vector of $A_\mu$. The state $B_\mu$ could be split into spin-$1/2$ and spin-$3/2$ degenerate states in the heavy quark limit corresponding to $\Sigma_Q$ and $\Sigma_Q^{\ast}$, respectively \cite{Mannel:1991ii,Roberts:1992xm,Falk:1991nq,Guo:1998ef}. According to Refs.~\cite{Mannel:1991ii,Roberts:1992xm,Hussain:1993qd}, the doublet $\{\Sigma_Q, \Sigma_Q^\ast\}$ could be represented by $B_\mu^m$, where $m=1$ $(2)$ stands for spin-$1/2$ $(3/2)$. The explicit forms of $B_\mu^1$ and $B_\mu^2$ are \cite{Falk:1991nq}
\begin{equation}\label{baryon}
  B_\mu^1(v) = \frac{1}{\sqrt 3}(\gamma_\mu+v_\mu)\gamma_5u(v),
\end{equation}
\begin{equation}\label{baryon2}
  B_\mu^2(v) = u_\mu(v),
\end{equation}
where $v$ is the velocity of $\Sigma_Q^{(\ast)}$, $u(v)$ is the Dirac spinor and $u_\mu(v)$ is the Rarita-Schwinger vector spinor \cite{Falk:1991nq}. Besides, $B_\mu^m(v)$ satisfies the conditions \cite{Guo:1998ef}:
\begin{equation}
  v\!\!\!/B^m_\mu=B^m_\mu(v),\quad\quad\quad v^\mu B^m_\mu(v)=0,\quad\quad\quad \gamma^\mu B^2_\mu=0.
\end{equation}

The BS wave function of $\Sigma_Q^{(\ast)}$ is defined as
\begin{equation}\label{Sigma}
  \chi_\mu(x_1,x_2,P)=\langle0|T\psi_Q(x_1)A_\mu(x_2)|\Sigma_Q^{(\ast)}(P)\rangle.
\end{equation}
Define $\eta_1=\frac{m_Q}{m_Q+m_A}$, $\eta_2=\frac{m_A}{m_Q+m_A}$, where $m_Q$ and $m_A$ denote the masses of the heavy quark and the axial-vector diquark, respectively. Then, $\Sigma_Q^{(\ast)}$ follows the similar BS equation to $\Lambda_Q$:
\begin{equation}\label{bsforsigma}
  \chi^\mu_{Pm}(p)=S_F(p_1)\int\frac{d^4q}{(2\pi)^4}G_{\rho\nu}(P,p,q)\chi_{Pm}^\nu(q)S^{\mu\rho}_D(-p_2),
\end{equation}
where $\chi_{Pm}^\mu$ is the BS wave function of $\Sigma_Q^{(\ast)}$ in momentum space, $G_{\rho\nu}(P,p,q)$ is the kernel in the $\Sigma_Q^{(\ast)}$ case, which is similar to the form used in the $\Lambda_Q$ case \cite{Guo:1996jj,Guo:1998ef}, $S_F(p_1)$ and $S_D^{\mu\rho}$ are the propagators of the heavy quark and the axial-vector diquark, respectively.

Due to the flavor and spin symmetries in the heavy quark limit, and taking into account Lorentz transformation, P-parity transformation and spin rotation, we have the following form for the BS wave function of $\Sigma_Q^{(\ast)}$ \cite{Falk:1991nq,Dai:1993kp,Guo:1998ef}:
\begin{equation}\label{realexpansion2}
  \chi^\mu_{Pm}=AB^\mu_m(v)+Cv^\mu p_{t\nu} B^\nu_m(v)+Dp^\mu_t p_{t\nu} B^\nu_m(v),
\end{equation}
where $A$, $C$ and $D$ are Lorentz scalar functions. After substituting the explicit forms of the propagators and the kernel $G_{\rho\nu}(P,p,q)$ into Eq.~\eqref{bsforsigma}, we obtain three coupled integral equations,
\begin{align}\label{A}
  \widetilde A(p_t^2)&=-\frac{1}{2w_p(\eta_1m_{\Sigma_Q^{(\ast)}}-\omega_Q-w_p)}\int\frac{d^3q_t}{(2\pi)^3}\left[\widetilde A(q_t^2)(\widetilde V_1-2w_p\widetilde V_2)\right. \nonumber\\
                     &-\widetilde C(q_t^2)\widetilde V_2\frac{p_t^2q_t^2-(p_t\cdot q_t)^2}{2p_t^2}+\widetilde D(q_t^2)(\widetilde V_1-2w_p\widetilde V_2)\frac{p_t^2q_t^2-(p_t\cdot q_t)^2}{2p_t^2}\Big],
\end{align}
\begin{align}\label{C}
  \widetilde C(p_t^2)&=-\frac{1}{2m_D^2w_p(\eta_1m_{\Sigma_Q^{(\ast)}}-\omega_Q-w_p)}
  \int\frac{d^3q_t}{(2\pi)^3}\left\{\widetilde A(q_t^2)\left[w_p\widetilde V_1\right.\right.\nonumber\\&-(m_D^2-w_p(\eta_1m_{\Sigma_Q^{(\ast)}}-\omega_Q))\widetilde V_2\Big] \nonumber\\
  &+\widetilde C(q_t^2)\left[\frac{(p_t\cdot q_t)}{p_t^2}(m_D^2-w_p(\eta_1m_{\Sigma_Q^{(\ast)}}-\omega_Q))\widetilde V_1-\frac{(p_t\cdot q_t)^2}{p_t^2}w_p\widetilde V_2\right] \nonumber\\
  &+\widetilde D(q_t^2)\left[\frac{(p_t\cdot q_t)^2}{p_t^2}w_p\widetilde V_1-\frac{(p_t\cdot q_t)^2}{p_t^2}(m_D^2+w_p(\eta_1m_{\Sigma_Q^{(\ast)}}-\omega_Q))\widetilde V_2\right]\Big\},
\end{align}
\begin{align}\label{D}
  \widetilde D(p_t^2)&=\frac{1}{2m_D^2w_p(\eta_1m_{\Sigma_Q^{(\ast)}}-\omega_Q-w_p)}\int\frac{d^3q_t}{(2\pi)^3}\left\{\widetilde A(q_t^2)(\widetilde V_1-w_p\widetilde V_2)\right. \nonumber\\
  &+\widetilde C(q_t^2)\left[-\frac{p_t\cdot q_t}{p_t^2}w_p\widetilde V_1+\frac{m_D^2(3(p_t\cdot q_t)^2-p_t^2q_t^2)-2p_t^2(p_t\cdot q_t)^2}{2p_t^4}\widetilde V_2\right]\nonumber\\
  &+\widetilde D(q_t^2)\left[\frac{-m_D^2(3(p_t\cdot q_t)^2-p_t^2q_t^2)+2p_t^2(p_t\cdot q_t)^2}{2p_t^4}\widetilde V_1\right.\nonumber\\&-\frac{p_t^2(p_t\cdot q_t)^2-m_D^2(3(p_t\cdot q_t)^2-p_t^2q_t^2)}{p_t^4}w_p\widetilde V_2\Big]\Big\},
\end{align}
where $w_p=\sqrt{m_A^2-p_t^2}$, and terms like $\int\frac{d^4q}{(2\pi)^4}q^\mu f$ and $\int\frac{d^4q}{(2\pi)^4}q_t^\mu q_t^\nu f$ can be expaned to terms including $p_t^\mu$ and $v^\mu$ on the grounds of Lorentz invariance \cite{Guo:1998ef}.

For the normalization of the BS wave function of $\Sigma_Q^{(\ast)}$, we can use the similar technique used in the $\Lambda_Q$ case. The Isgur-Wise function for $\Sigma_Q^{(\ast)}$ has the following relation with the transition matrix $\Sigma_b^{(\ast)}\to\Sigma_c^{(\ast)}$ at the zero recoil point:
\begin{equation}\label{wise}
  \langle\Sigma_c^{(\ast)}(v)|\bar c\Gamma b|\Sigma_b^{(\ast)}(v)\rangle=\xi(1)\bar B_{m^\prime\mu}(v)\Gamma B^\mu_m(v),
\end{equation}
where $\xi(1)$ is the Isgur-Wise function for $\Sigma_Q^{(\ast)}$ at the zero recoil point. Meanwhile, the above transition matrix can be expressed with the BS wave function of $\Sigma_Q^{(\ast)}$ as the following:
\begin{equation}\label{link}
   \langle\Sigma_c^{(\ast)}(v)|\bar c\Gamma b|\Sigma_b^{(\ast)}(v)\rangle=\int\frac{d^4p}{(2\pi)^4}\frac{i}{\eta_1m_{\Sigma_Q^{(\ast)}}-\omega_Q+p_l+i\varepsilon}\bar\chi^\mu_{P m^\prime}(p)\Gamma\int\frac{d^4q}{(2\pi)^4}G_{\mu\nu}(P,p,q)\chi^\nu_{Pm}(p),
\end{equation}
substituting Eq.~\eqref{link} into \eqref{wise}, we arrive at
\begin{align}\label{xi2}
  \xi(1)=&\int\frac{p_t^2dp_t}{4\pi^2}\frac{2}{\eta_1m_{\Sigma_Q^{(\ast)}}-\omega_Q-w_p}\left[\widetilde A(p_t^2)h_1(|p_t|)-\frac{1}{3}p_t^2\widetilde C(p_t^2)h_2(|p_t|)\right.\nonumber\\
  &-\frac{1}{3}p_t^2\widetilde D(p_t^2)h_3(|p_t|)+\frac{1}{6m_A^2}p_t^2h_2(|p_t|)h_4(|p_t|)\left.\right],
\end{align}
where $h_i(|p_t|)(i=1,2,3,4)$ are of the forms:
\begin{align}\label{h1}
  h_1(|p_t|)&=\int\frac{q_t^2dq_t}{4\pi^2}\left\{8\pi\kappa F_1[\widetilde A(q_t^2)-\frac{1}{3}q_t^2\widetilde D(q_t^2)]+F_2\left[2w_p\widetilde A(q_t^2)-\frac{1}{3}q_t^2\widetilde C(q_t^2)\right.\right. \nonumber\\
  &-\frac{2}{3}w_pq_t^2\widetilde D(q_t^2)\left.\right]-8\pi\kappa F_1[\widetilde A(p_t^2)-\frac{1}{3}p_t^2\widetilde D(p_t^2)]\left.\right\},
\end{align}
\begin{equation}\label{h2}
  h_2(|p_t|)=\int\frac{q_t^2dq_t}{4\pi^2}\left[F_2\widetilde A(q_t^2)+8\pi\kappa F_3\frac{1}{p_t^2}\widetilde C(q_t^2)
  +F_4\frac{1}{p_t^2}\widetilde D(q_t^2)-8\pi\kappa F_1\widetilde C(p_t^2)\right],
\end{equation}
\begin{align}\label{h3}
  h_3(|p_t|)&=\int\frac{q_t^2dq_t}{4\pi^2}\left\{8\pi\kappa F_1\widetilde A(q_t^2)+2w_pF_2\widetilde A(q_t^2)-8\pi\kappa F_5\frac{1}{p_t^2}\widetilde D(q_t^2)\right.\nonumber\\
  &+F_4\frac{1}{p_t^2}[\widetilde C(q_t^2+2w_p\widetilde D(q_t^2))]-8\pi\kappa F_1[p_t^2\widetilde D(p_t^2)-\widetilde A(p_t^2)]\left.\right\},
\end{align}
\begin{equation}\label{h4}
  h_4(|p_t|)=\int\frac{q_t^2dq_t}{4\pi^2}\left\{-F_2\widetilde A(q_t^2)+8\pi\kappa F_3\frac{1}{p_t^2}\widetilde C(q_t^2)-F_4\frac{1}{p_t^2}\widetilde D(q_t^2)-8\pi\kappa F_1\widetilde C(p_t^2)\right\},
\end{equation}
where $F_i(i=1,2,3,4,5)$ are functions of $|p_t|$ and $|q_t|$, the explicit forms of which can be found in Ref.~\cite{Guo:1998ef}, $\beta=\alpha_{seff}^2Q_0^2$. In this way, scalar functions $\widetilde A(p_t^2)$, $\widetilde C(p_t^2)$ and $\widetilde D(p_t^2)$ can be normalized with the aid of Eqs.~\eqref{xi2}-\eqref{h4}.
\section{Calculations of residues for $\Lambda_Q$-type and $\Sigma_Q$-type baryons}
\subsection{The numerical solutions for the BS wave functions of $\Lambda_Q$-type and $\Sigma_Q$-type baryons}
In this section, we will solve the BS equations numerically and then apply the results to calculate the residues of $\Lambda_Q$-type and $\Sigma_Q$-type baryons, which are defined in Eqs.~\eqref{def1}-\eqref{def3}. To solve Eq.~\eqref{bsforlambda} and the three coupled integral equations \eqref{A}, \eqref{C} and \eqref{D} numerically, we have to discretize the integration regions into $n$ pieces with $n$ being sufficiently large. Then the equations become eigenvalue equations and can be solved with Gaussian integration method.

There are several parameters in our model for $\Lambda_Q$-type and $\Sigma_Q$-type baryons: $\alpha_{seff}$, $\kappa$, $Q_0^2$, $m_D$, $m_A$. The coupling parameter $\kappa$ is taken to in the range $0.02\,\rm GeV^3\sim0.08\,\rm GeV^3$ in the baryon case \cite{Guo:2007tn}. Once $\kappa$ is settled, $\alpha_{seff}$ can be determined by solving the eigenvalue equations because of the constraint between $\kappa$ and $\alpha_{seff}$. $Q_0^2=3.2\,\rm GeV^2$ extracted from the electromagnetic form factor of the proton is taken in both $\Lambda_Q$-type and $\Sigma_Q$-type cases, as it is associated with the light diquark \cite{Guo:1998ef,Guo:1996jj,Guo:2007tn,Guo:1999ss}. For the heavy quark masses, we will take the values that have been used in many papers \cite{Guo:1996jj,Guo:1994fn,Dai:1993np}: $m_b=5.02\,\rm GeV$, $m_c=1.58\,\rm GeV$, because the predictions they lead are consistent with experiment data. Besides, we will let the diquark masses vary in reasonable regions: $0.60\sim0.80\,\rm GeV$ for the light scalar diquark $ud$ \cite{Liu:2018tqe,Guo:1996jj}, $0.90\sim1.00\,\rm GeV$ for the light scalar diquark $us$, $0.90\sim1.10\,\rm GeV$ for the light axial-vector diqaurk $uu$, $ud$ or $dd$, $1.05\sim1.15\,\rm GeV$ for the axial-vector diquark $us$ or $ds$ and $1.15\sim1.25\,\rm GeV$ for the axial-vector diquark $ss$. On top of that, the parameter $\mu$, which is introuduced here to cancel the infrared divergence, will be taken to approach $0$ in the end of calculations.

For the masses of $\Lambda_Q$-type and $\Sigma_Q$-type baryons, we have the following expressions in the leading order of $1/m_Q$ expansion:
\begin{eqnarray}\label{mexpansion}
  m_i&=&m_Q+m_{D_i}+E_i\quad\quad(i=\Lambda_Q,\Xi_Q^A),\\
  m_i&=&m_Q+m_{A_i}+E_i\quad\quad(i=\Sigma_Q^{(\ast)},\Xi_Q^{S(\ast)},\Omega_Q^{(\ast)}),
\end{eqnarray}
where $E_i$ is the binding energy. It can be seen that the parameters $m_{D_i(A_i)}$ and $E_i$ are constrained by the relation $m_{D_i(A_i)}+E_i=m_i-m_Q$, where we omit the $1/m_Q$ corrections. Thus, we will have $m_{D_i}+E_i$ as $0.60\,\rm GeV$ for $i=\Lambda_Q$ and $0.77\,\rm GeV$ for $i=\Xi_Q^A$, and $m_{A_i}+E_i$ as $0.79\,\rm GeV$ for $i=\Sigma_Q^{(\ast)}$, $0.93\,\rm GeV$ for $i=\Xi_Q^{S(\ast)}$ and $1.03\,\rm GeV$ for $i=\Omega_Q^{(\ast)}$ \cite{Patrignani:2016xqp}. In Table~\ref{LMN}, we give the values of $\alpha_{seff}$ we obtain while solving the BS equation for $m_D=0.60,0.70,0.80\,\rm GeV$ with different values of $\kappa$ ranging from $0.02\,\rm GeV^3$ to $0.08\,\rm GeV^3$ for $\Lambda_Q$ \cite{Liu:2018tqe}.
\begin{table*}[h!]
\renewcommand\arraystretch{0.9}
\centering
\caption{\vadjust{\vspace{-5pt}}The values of $\alpha_{seff}$ for $\Lambda_Q$ with different values of $\kappa$ and $m_D$ under the heavy quark limit.}\label{LMN}\vspace{1.2mm}
\begin{tabular*}{\textwidth}{@{\extracolsep{\fill}}ccccc}
\hline
 $\kappa(\,\rm GeV^3)$    &0.02&0.04&0.06&0.08 \\
\hline
 $m_D=0.60\,\rm GeV$    &0.58&0.65&0.68&0.71 \\
 $m_D=0.70\,\rm GeV$    &0.69&0.72&0.75&0.77 \\
 $m_D=0.80\,\rm GeV$    &0.77&0.79&0.81&0.82 \\
\hline
\hline
\end{tabular*}
\end{table*}
\begin{table*}[h!]
\renewcommand\arraystretch{0.9}
\centering
\caption{\vadjust{\vspace{-5pt}}The values of $\alpha_{seff}$ for $\Xi_Q^A$ with different values of $\kappa$ and $m_D$ under the heavy quark limit.}\label{LMN3}\vspace{1.2mm}
\begin{tabular*}{\textwidth}{@{\extracolsep{\fill}}ccccc}
\hline
 $\kappa(\,\rm GeV^3)$    &0.02&0.04&0.06&0.08 \\
\hline
 $m_D=0.90\,\rm GeV$    &0.69&0.71&0.73&0.74 \\
 $m_D=0.95\,\rm GeV$    &0.73&0.75&0.76&0.77 \\
 $m_D=1.00\,\rm GeV$    &0.77&0.78&0.79&0.80 \\
\hline
\hline
\end{tabular*}
\end{table*}
As we can see from the data in Table~\ref{LMN}, once the parameters $m_D$ and $Q_0^2$ are fixed, $\alpha_{seff}$ for $\Lambda_Q$ will undergo some notable change with the increase of $\kappa$, and for each $\kappa$, $\alpha_{seff}$ also increases as $m_D$ increases. When the confinement parameter $\kappa$ is relatively small, the value of $m_D$ will impose bigger influence on $\alpha_{seff}$. On top of that, we also find that when $\kappa$ and the mass for diqaurk we choose
are small enough, there will not be numerical solution for the BS wave function of $\Lambda_Q$, which suggests the value of $m_D$ should be at least $0.60\,\rm GeV$. In general, $\alpha_{seff}$ in $\Xi_Q^A$ follows a similar pattern to the change of $\alpha_{seff}$ in $\Lambda_Q$, but the increase in $\alpha_{seff}$ for $\Xi_Q^A$ as $\kappa$ increases is relatively small compared to the case in $\Lambda_Q$ when the mass of diquark is small.

For the case of $\Sigma_Q$-type baryons, we have to solve three coupled integral equations \eqref{A}, \eqref{C} and \eqref{D} to obtain the numerical solution of $\widetilde A(p_t^2)$, $\widetilde C(p_t^2)$ and $\widetilde D(p_t^2)$. After discretization, these three equations can be changed to three matrix equations,
\begin{equation}\label{AA}
  \widetilde A=Z_1\widetilde A+Z_2\widetilde C+Z_3\widetilde D,
\end{equation}
\begin{equation}\label{CC}
  0=R_1\widetilde A+R_2\widetilde C+R_3\widetilde D,
\end{equation}
\begin{equation}\label{DD}
  0=T_1\widetilde A+T_2\widetilde C+T_3\widetilde D,
\end{equation}
where $Z_i$, $R_i$ and $T_i$ $(i=1,2,3)$ are $n\times n$ matrices. With the help of Eqs.~\eqref{CC} and \eqref{DD}, $\widetilde C$ and $\widetilde D$ can be eliminated from Eq.~\eqref{AA}. Thus, we can solve the wave function $\widetilde A$ first and then we can obtain $\widetilde C$ and $\widetilde D$ by using the same technique. In Table~\ref{LMN2}, we give the values of $\alpha_{seff}$ for $\Sigma_Q^{(\ast)}$ in the ranges of the parameters in our model.

As it is shown in Table~\ref{LMN2} that there appear small changes in the values of $\alpha_{seff}$ as $\kappa$ varies. We find no numerical solutions for $\alpha_{seff}$ when the mass of the axial-vector diquark is $1.10\,\rm GeV$, which suggests that the maximum value for the mass of the light axial-vector diquark in $\Sigma_Q^{(\ast)}$ should be below $1.10\,\rm GeV$, and we find the maximum mass value for $m_A$ allowed in our model is $1.06\,\rm GeV$.
\begin{table*}[h!]
\renewcommand\arraystretch{0.9}
\centering
\caption{\vadjust{\vspace{-5pt}}The values of $\alpha_{seff}$ for $\Sigma_Q^{(\ast)}$ with different values of $\kappa$ and $m_D$ under the heavy quark limit.}\vspace{1.2mm}\label{LMN2}
\begin{tabular*}{\textwidth}{@{\extracolsep{\fill}}cccccc}
\hline
 $\kappa(\rm GeV^3)$    &0.02&0.04&0.06&0.08 \\
\hline
 $m_A=0.90\,\rm GeV$    &0.62&0.63&0.63&0.64 \\
 $m_A=1.00\,\rm GeV$    &0.69&0.70&0.70&0.73 \\
 $m_A=1.06\,\rm GeV$    &0.75&0.76&-&- \\
\hline
\hline
\end{tabular*}
\end{table*}
\begin{table*}[h!]
\renewcommand\arraystretch{0.9}
\centering
\caption{\vadjust{\vspace{-5pt}}The values of $\alpha_{seff}$ for $\Xi_Q^{S(\ast)}$ with different values of $\kappa$ and $m_D$ under the heavy quark limit.}\vspace{1.2mm}\label{LMN5}
\begin{tabular*}{\textwidth}{@{\extracolsep{\fill}}cccccc}
\hline
 $\kappa(\rm GeV^3)$    &0.02&0.04&0.06&0.08 \\
\hline
 $m_A=1.05\,\rm GeV$    &0.64&0.65&0.66&0.67 \\
 $m_A=1.10\,\rm GeV$    &0.69&0.69&0.70&0.71 \\
 $m_A=1.15\,\rm GeV$    &0.72&0.73&0.73&0.74 \\
\hline
\hline
\end{tabular*}
\end{table*}
\begin{table*}[h!]
\renewcommand\arraystretch{0.9}
\centering
\caption{\vadjust{\vspace{-5pt}}The values of $\alpha_{seff}$ for $\Omega_Q^{(\ast)}$ with different values of $\kappa$ and $m_D$ under the heavy quark limit.}\vspace{1.2mm}\label{LMN4}
\begin{tabular*}{\textwidth}{@{\extracolsep{\fill}}cccccc}
\hline
 $\kappa(\rm GeV^3)$    &0.02&0.04&0.06&0.08 \\
\hline
 $m_A=1.15\,\rm GeV$    &0.58&0.60&0.63&0.64 \\
 $m_A=1.20\,\rm GeV$    &0.62&0.63&0.65&0.67 \\
 $m_A=1.25\,\rm GeV$    &0.65&0.67&0.67&0.69 \\
\hline
\hline
\end{tabular*}
\end{table*}

The values of $\alpha_{seff}$ are given in Table~\ref{LMN5} and \ref{LMN4} for $\Xi_Q{S(\ast)}$ and $\Omega_Q^{(\ast)}$ respectively, where we can see that $\alpha_{seff}$ for $\Xi_Q^{S(\ast)}$ and $\Omega_Q^{(\ast)}$ increases with the increase in $\kappa$ and the mass of diquark. Generally, the changes of $\alpha_{seff}$ in these three baryons ($\Sigma_Q^{(\ast)}$, $\Xi_Q^{S(\ast)}$ and $\Omega_Q^{(\ast)}$) with different values of $\kappa$ and $m_A$ are quite similar due to the approximate $SU(3)$ flavor symmetry.

\begin{figure}[h!]
\scalebox{0.9}[0.9]{\includegraphics{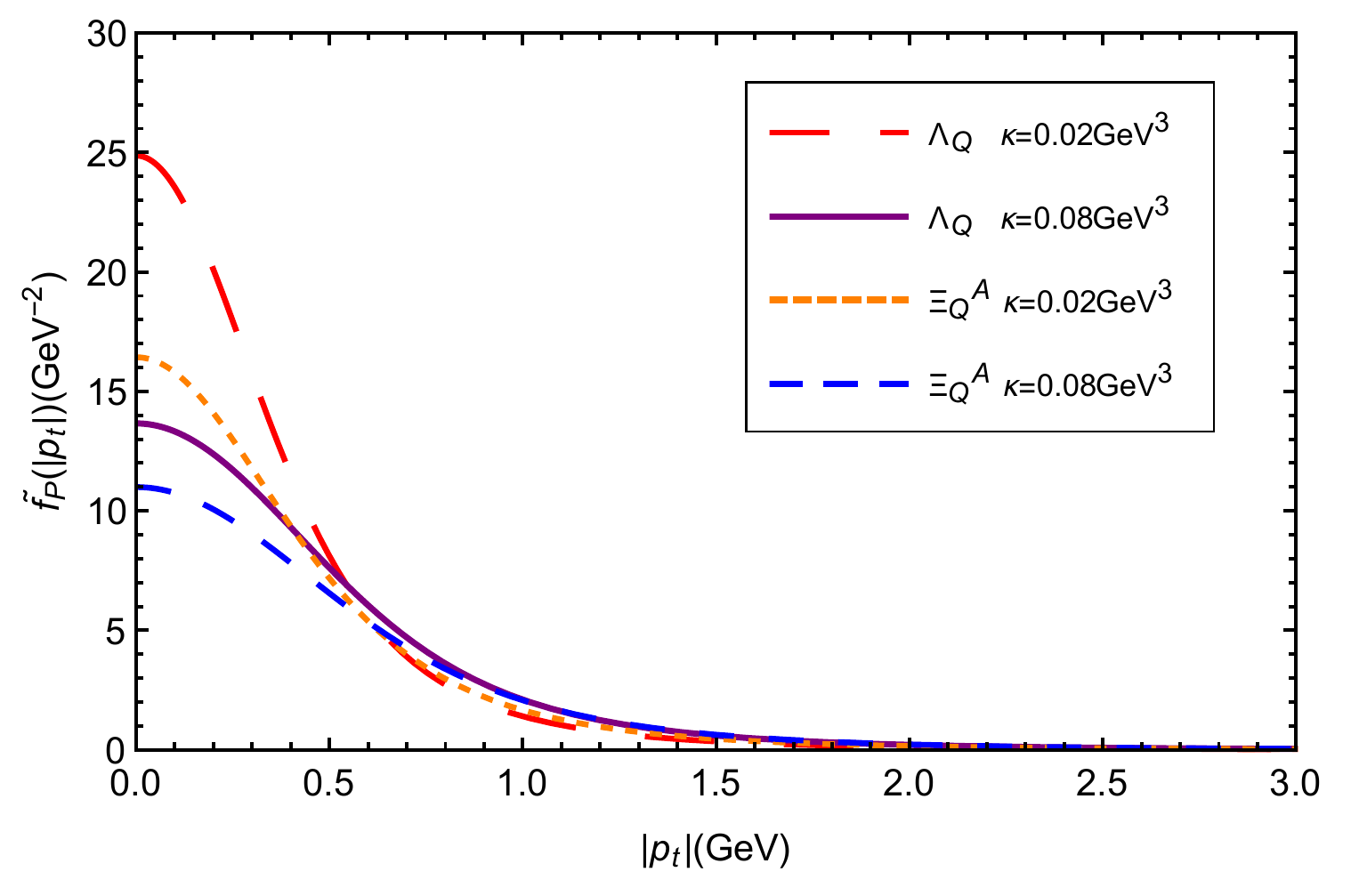}}
\caption{The normalized BS wave functions for $\Lambda_Q$-type baryons, $\Lambda_Q$ ($m_D=0.70\,\rm GeV$) and $\Xi_Q^A$ ($m_D=0.95\,\rm GeV$), with $\kappa=0.02\,\rm GeV^3$ and $\kappa=0.08\,\rm GeV^3$ respectively.}\label{fig2}
\centering
\end{figure}

\begin{figure}[h!]
\scalebox{0.9}[0.9]{\includegraphics{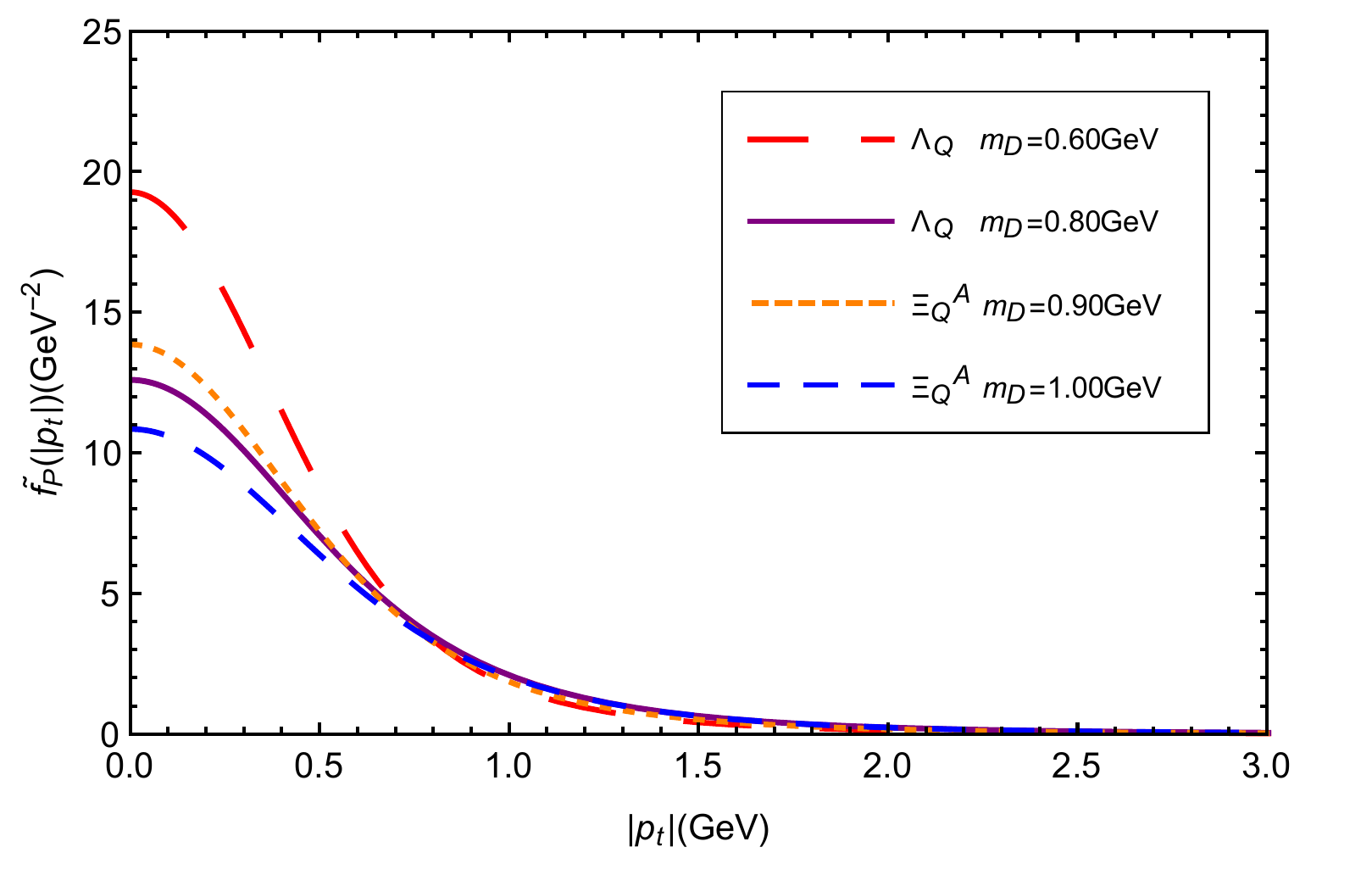}}
\caption{The normalized BS wave functions for $\Lambda_Q$-type baryons, $\Lambda_Q$ ($m_D=0.60\sim0.80\,\rm GeV$) and $\Xi_Q^A$ ($m_D=0.90\sim1.00\,\rm GeV$) when $\kappa=0.06\,\rm GeV^3$ for both baryons.}\label{fig3}
\centering
\end{figure}

In Fig.~\ref{fig2}, we plot the BS wave function for $\Lambda_Q$-type baryons with respect to $|p_t|$. It can be seen from these curves that the shapes of all these BS wave functions are similar, and clearly, the BS wave functions for $\Lambda_Q$ are steeper than those for $\Xi_Q^A$, which can be attributed to the more heavy diqaurk in $\Xi_Q^A$. Similarly, we obtain four BS wave functions for $\Lambda_Q$ and $\Xi_Q^A$ in different scenarios in Fig.~\ref{fig3}, where we have $\kappa$ fixed at $0.06\,\rm GeV^3$, and allow the mass of diquark of $\Lambda_Q$ and $\Xi_Q^A$ vary in their regions. It can be seen the results are similar to Fig.~\ref{fig2} and Fig.~\ref{fig3}.
\begin{figure}[h!]
\scalebox{0.9}[0.9]{\includegraphics{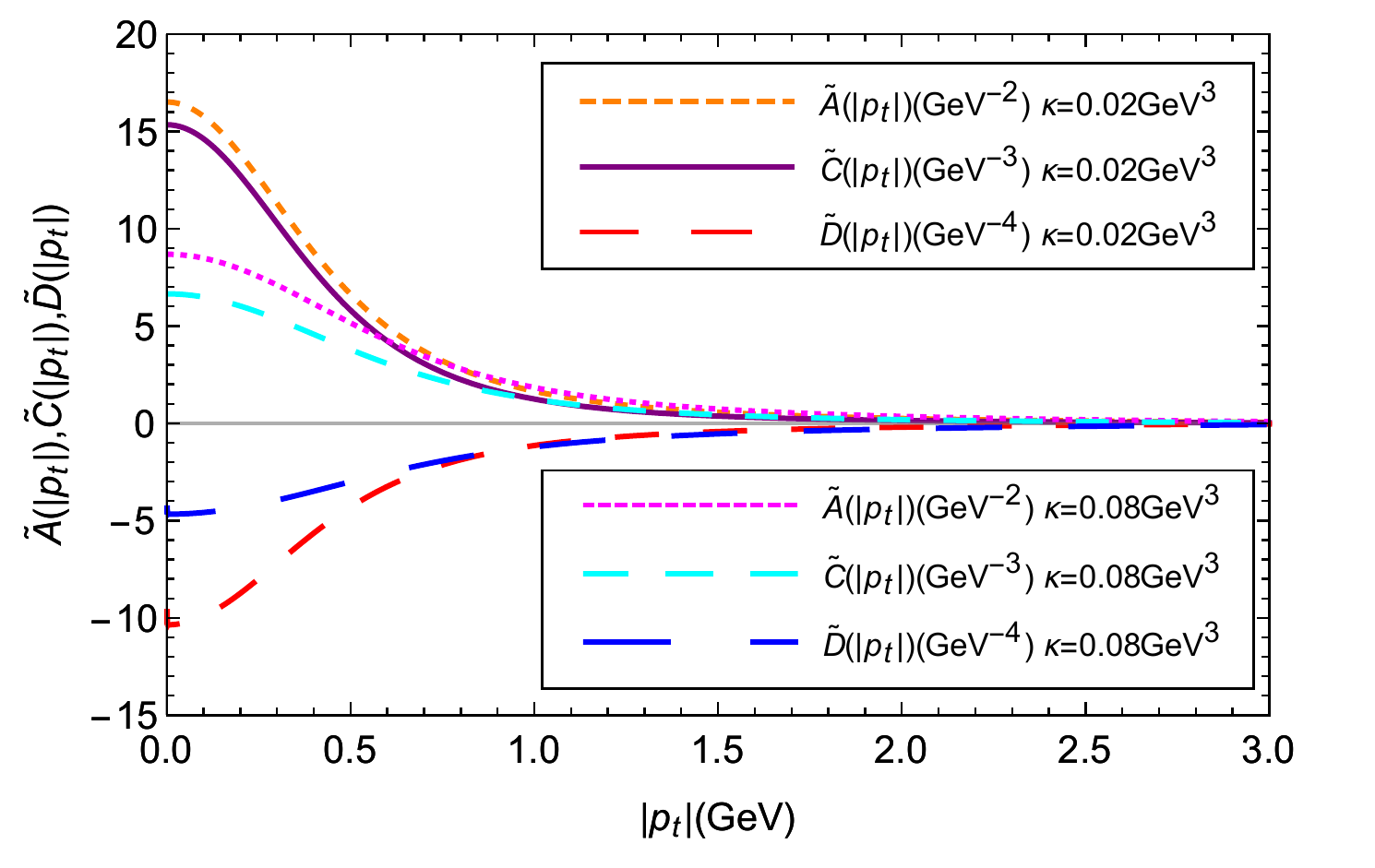}}
\caption{The normalized BS wave functions of $\Sigma_Q^{(\ast)}$ with $\kappa=0.02\,\rm GeV^3$ and $\kappa=0.08\,\rm GeV^3$ when $m_A=0.90\,\rm GeV$.}\label{fig4}
\centering
\end{figure}
\begin{figure}[h!]
\scalebox{0.9}[0.9]{\includegraphics{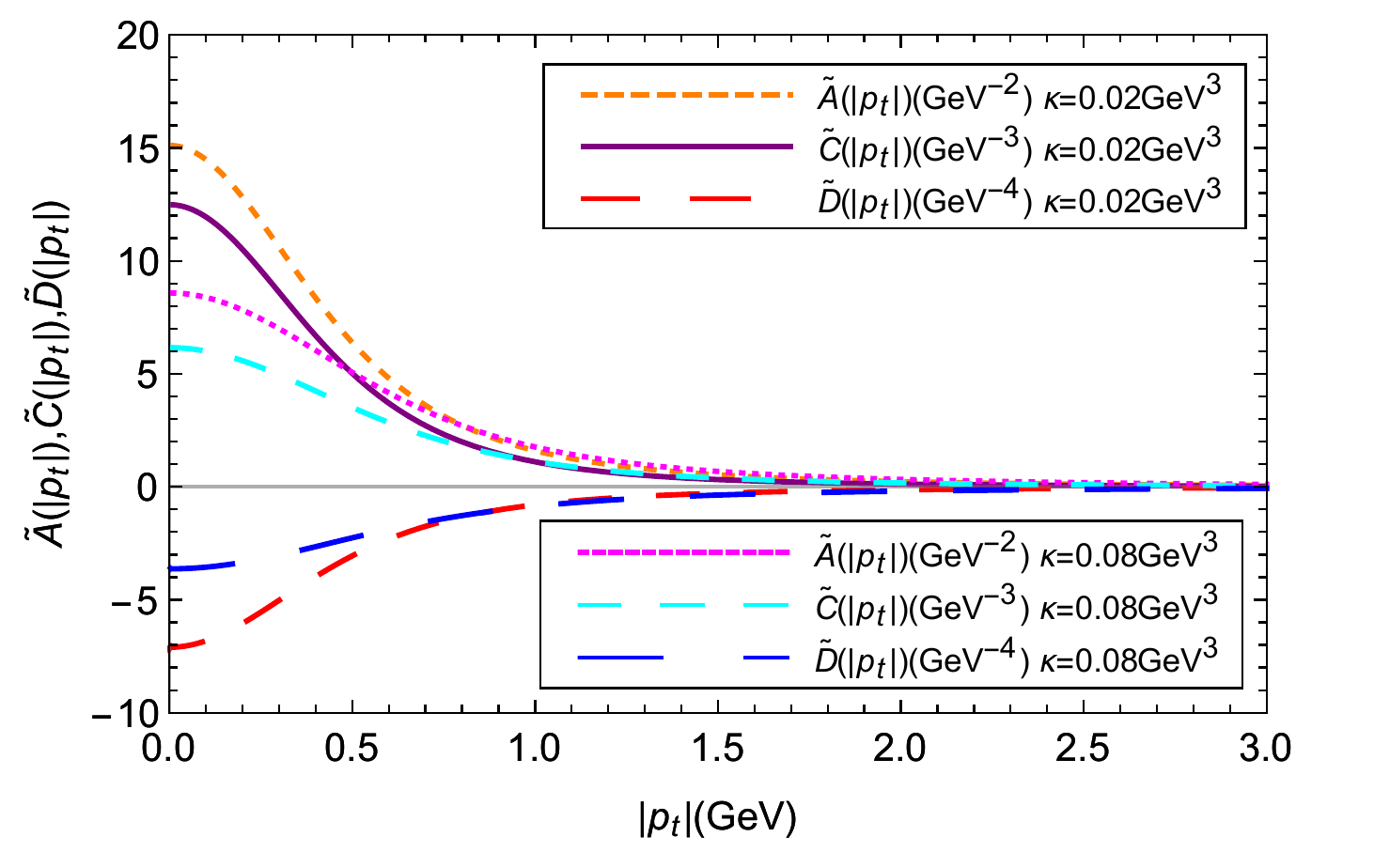}}
\caption{The normalized BS wave functions of $\Xi_Q^{S(\ast)}$ with $\kappa=0.02\,\rm GeV^3$ and $\kappa=0.08\,\rm GeV^3$ when $m_A=1.05\,\rm GeV$.}\label{fig5}
\centering
\end{figure}
\begin{figure}[h!]
\scalebox{0.9}[0.9]{\includegraphics{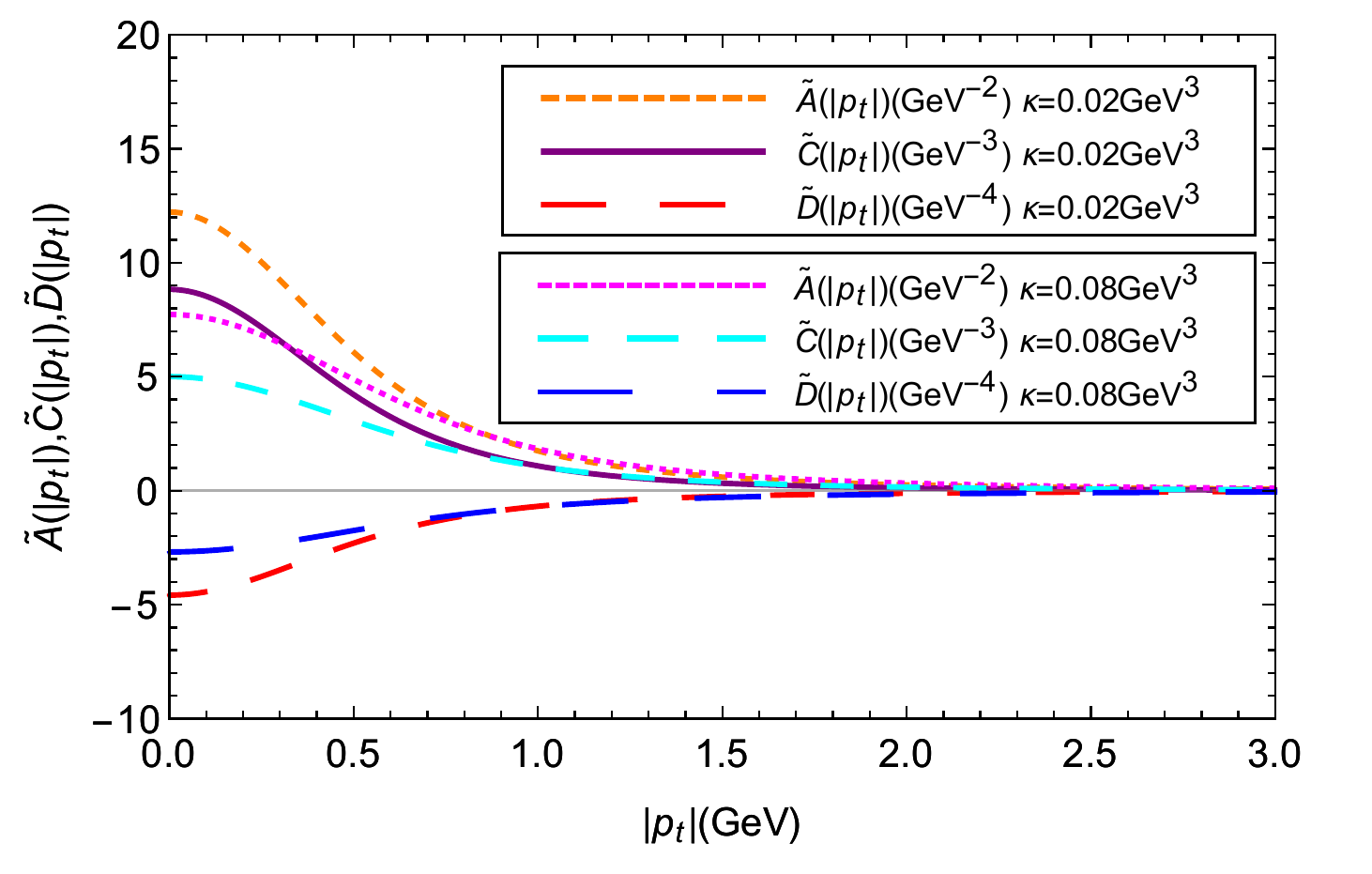}}
\caption{The normalized BS wave functions of $\Omega_Q^{(\ast)}$ with $\kappa=0.02\,\rm GeV^3$ and $\kappa=0.08\,\rm GeV^3$ when $m_A=1.20\,\rm GeV$.}\label{fig6}
\centering
\end{figure}

We present the solutions of the BS wave functions for $\Sigma_Q$-type baryon in Figs.~\ref{fig4}, \ref{fig5} and \ref{fig6}, where we give the numerical results of BS wave functions for $\Sigma_Q^{(\ast)}$, $\Xi_Q^{S(\ast)}$ and $\Omega_Q^{(\ast)}$, with $\kappa$ in the range $0.02\sim0.08\,\rm GeV^3$ for all the $\Sigma_Q$-type baryons in this work. For example, the BS wave function of $\Sigma_Q^{(\ast)}$ is solved with different values of $\kappa$ when $m_D=1.00\,\rm GeV$. In comparison, we also illustrate the other two $\Sigma_Q$-type baryons, we can see that curves are becoming smoother and wider as the mass $m_A$ increases. Besides, it can be easily seen from these figures that the BS wave functions for $\Sigma_Q^{(\ast)}$ follow the same pattern as those for $\Lambda_Q$.
\subsection{Calculations of residues of $\Lambda_Q$-type and $\Sigma_Q$-type baryons}
From the baryonic currents for $\Lambda_Q$-type and $\Sigma_Q$-type baryons constructed in the quark-diquark model in Sec.~\uppercase\expandafter{\romannumeral2}, we can obtain the relations between the residues and the BS wave functions for $\Lambda_Q$-type and $\Sigma_Q$-type baryons as follows:
\begin{equation}\label{la}
  \langle 0|J_{\Lambda_Q}|\Lambda_Q\rangle= \langle 0|D\psi|\Lambda_Q\rangle=\int\frac{d^4p}{(2\pi)^4}\chi_{P}(p),
\end{equation}
\begin{equation}\label{si}
  \langle 0|J_{\Sigma_Q}|\Sigma_Q\rangle= \langle 0|A^\mu\gamma_\mu\gamma_5\psi|\Sigma_Q\rangle=\int\frac{d^4p}{(2\pi)^4}\gamma_\mu\gamma_5\chi_{Pm}^\mu(p)\quad\quad(m=1).
\end{equation}
Explicitly, the residues of $\Lambda_Q$-type baryons can be expressed as (take $\Lambda_Q$ as the example)
\begin{equation}\label{residue1}
  f_{\Lambda_Q}=\int\frac{dp_t}{2\pi^2}\widetilde f_P(|p_t|)p_t^2.
\end{equation}
For $\Sigma_Q$-type baryons, we have $B^\mu_1=\frac{1}{\sqrt 3}(\gamma^\mu+v^\mu)\gamma^5u(v)$, substituting this equation into Eq.~\eqref{si}, we arrive at
\begin{equation}\label{residuess}
  f_{\Sigma_Q}u_{\Sigma_Q}=\int\frac{d^4p}{(2\pi)^4}\frac{1}{\sqrt 3}\left[A(4-v\!\!\!/)-Cv\!\!\!/\gamma_5p_t\!\!\!\!/\gamma_5-Dp_t\!\!\!\!/\gamma_5p_t\!\!\!\!/\gamma_5\right]u_{\Sigma_Q},
\end{equation}
since the Dirac spinor $u_{\Sigma_Q}$ satisfies the condition: $v\!\!\!/u_{\Sigma_Q}=u_{\Sigma_Q}$, and $\int\frac{d^4p}{(2\pi)^4}p_t^\mu f(|p|^2)=0$ ($f(|p^2|)$ stands for a function of $|p|^2$), we have
\begin{equation}\label{residuesss}
  f_{\Sigma_Q}=\frac{1}{\sqrt 3}\int\frac{d^4p}{(2\pi)^4}\left[3A(p)-|p_t|^2D(p)\right].
\end{equation}
Finally, completing the integral $\int\frac{dp_l}{2\pi}$ in \eqref{residuesss} and integrating $\theta$ and $\phi$ in the spherical coordinate system, one immediately obtains
\begin{equation}\label{final2}
  f_{\Sigma_Q}=\frac{1}{\sqrt 3}\int\frac{dp_t}{2\pi^2}|p_t|^2\left[3\widetilde A(|p_t|^2)-|p_t|^2\widetilde D(|p_t|^2)\right].
\end{equation}
For $\Sigma_Q^{\ast}$-type baryons, it can be shown that in the leading order of HQET, we have (take $\Sigma_Q^\ast$ as the example)
\begin{equation}\label{ww}
  \langle0|J^\mu_{\Sigma_Q^\ast}|\Sigma_Q^\ast\rangle=\frac{1}{\sqrt3}f_{\Sigma_Q^\ast}u^\mu_{\Sigma_Q^\ast},
\end{equation}
substituting the explicit form of $J_{\Sigma_Q^\ast}$ into Eq.~\eqref{ww} and multiplying $\sqrt{3}p_{t\mu}$ on both sides of it, we have
\begin{equation}\label{www}
  \int\frac{d^4p}{(2\pi)^4}\sqrt3\left[A(p)+\frac{1}{3}D(p)p_t^2+\frac{1}{3}C(p)p\!\!\!/_t\right]=f_{\Sigma_Q^\ast},
\end{equation}
following the same routine used in the calculation of $f_{\Sigma_Q}$, we obtain
\begin{equation}\label{wwww}
  f_{\Sigma_Q^\ast}=\frac{1}{\sqrt 3}\int\frac{dp_t}{2\pi^2}|p_t|^2\left[3\widetilde A(|p_t|^2)-|p_t|^2\widetilde D(|p_t|^2)\right],
\end{equation}
so the residues of $\Sigma_Q^{\ast}$-type baryons are the same as those of $\Sigma_Q$-type baryons, which is consistent with the heavy quark symmetry.

In Table~\ref{res}, we give the numerical results for the residues of $\Lambda_Q$-type baryons $\Lambda_Q$ and $\Xi_Q^A$ with $m_D$ ranging from $0.60\,\rm GeV$ to $0.80\,\rm GeV$ and $0.90\,\rm GeV$ to $1.00\,\rm GeV$ respectively when $\kappa=0.02,0.04,0.06,0.08\,\rm GeV^3$. In general, we find that both $f_{\Lambda_Q}$ and $f_{\Xi_Q^A}$ increase as $m_D$ and $\kappa$ increase. Besides, it can be seen that the value of $f_{\Lambda_Q}$ changes from $0.103\,\rm GeV$ to $0.224\,\rm GeV$ in the variation ranges of the model parameters $\kappa$ and $m_D$, and $f_{\Xi_Q^A}$ changes from $0.143\,\rm GeV$ to $0.215\,\rm GeV$ with the variations of $\kappa$ and $m_D$. Furthermore, the dependence of the residue $f_{\Lambda_Q}$ on $m_D$ is stronger than that on $\kappa$. For example, the variation caused by the change of $m_D$ is about $0.079\,\rm GeV$ when $\kappa$ is $0.02\,\rm GeV^3$; by contrast, the variation is $0.042\,\rm GeV$ as $\kappa$ changes from $0.02\,\rm GeV^3$ to $0.08\,\rm GeV^3$ when $m_D$ is $0.80\,\rm GeV$. We find the same pattern for the dependence of $f_{\Xi_Q^A}$ on $\kappa$ and $m_D$. On top of that, we find that when $\kappa$ is fixed, the variation caused by the mass of diquark is larger for $\Lambda_Q$ than $\Xi_Q^A$, for example, when $\kappa=0.04\,\rm GeV^3$, the variation of the residue of $\Lambda_Q$ is $0.061\,\rm GeV$, but only $0.033\,\rm GeV$ for $\Xi_Q^A$.
\begin{table*}[h!]
\renewcommand\arraystretch{0.9}
\centering
\caption{\vadjust{\vspace{-5pt}}The residues of $\Lambda_Q$ and $\Xi_Q^A$ with different values of $\kappa$ and in the ranges of $m_D$ in our model under the heavy quark limit.}\label{res}
\begin{tabular*}{\textwidth}{@{\extracolsep{\fill}}cccccc}
\hline
 $\kappa(\rm GeV^3)$    &0.02&0.04&0.06&0.08 \\
\hline
 $f_{\Lambda_Q}(\rm GeV)$    &0.103$\sim$0.182&0.137$\sim$0.198&0.163$\sim$0.212&0.182$\sim$0.224 \\
\hline
 $f_{\Xi^A_Q}(\rm GeV)$      &0.143$\sim$0.181&0.161$\sim$0.194&0.176$\sim$0.205&0.189$\sim$0.215 \\
\hline
\hline
\end{tabular*}
\end{table*}

The results for the residues of $\Sigma_Q$-type baryons, $\Sigma_Q^{(\ast)}$, $\Xi_Q^{S(\ast)}$ and $\Omega_Q^{(\ast)}$, with different values of $\kappa$ and in the ranges of $m_A$ are given in Table~\ref{res2}. We find the residue of $\Sigma_Q^{(\ast)}$ is in the range $0.262\,\rm{GeV}\sim0.361\,\rm{GeV}$, with parameters $\kappa$ and $m_D$ varying in their regions. The residue of $\Xi_Q^{S(\ast)}$ is in the range $0.313\,\rm GeV\sim0.460\,\rm GeV$, which is generally larger than that of $\Sigma_Q^{(\ast)}$, and the residue of $\Omega_Q^{(\ast)}$ in the range $0.350\,\rm GeV\sim0.571\,\rm GeV$ are lager than those of both $\Sigma_Q^{(\ast)}$ and $\Xi_Q^{S(\ast)}$. The variations of the residues for $\Xi_Q^{S{(\ast)}}$ and $\Omega_Q^{(\ast)}$ are becoming smaller as $\kappa$ increases. Furthermore, when $\kappa$ is fixed, for example, at $0.04\,\rm GeV^3$, the variations of the residues increase as the masses of $\Sigma_Q$-type baryons increase. Besides, we find that the variation resulted by the mass of the diquark is wider for $\Xi_Q^{S(\ast)}$ than $\Sigma_Q^{(\ast)}$, for example, the variation is $0.026\,\rm GeV$ for $\Sigma_Q^{(\ast)}$, but $0.090\,\rm GeV$ for $\Xi_Q^{(\ast)}$ when $\kappa=0.02\,\rm GeV^3$. However, when $\kappa=0.08\,\rm GeV^3$, the variation of the residue is only $0.008\,\rm GeV$ for $\Xi_Q^{(\ast)}$, but $0.044\,\rm GeV$ for $\Sigma_Q^{(\ast)}$. It suggests that different values of $\kappa$ have different influences on the residues of $\Sigma_Q$-type baryons. It can be seen in general that the residues of $\Sigma_Q$-type baryons are larger than that of $\Lambda_Q$-type baryons, which is consistent with the residues from QCD sum rules \cite{Groote:1997ci}. The changes of residues of $\Sigma_Q$-type baryons with $\kappa$ and $m_A$ follow the similar pattern to that of $\Lambda_Q$-type baryons.
\begin{table*}[h!]
\renewcommand\arraystretch{0.9}
\centering
\caption{\vadjust{\vspace{-5pt}}The residues of $\Sigma_Q^{(\ast)}$, $\Xi_Q^{S(\ast)}$ and $\Omega_Q^{(\ast)}$ with different values of $\kappa$ and in the ranges of $m_A$ under the heavy quark limit.}\label{res2}\vspace{0.1cm}
\begin{tabular*}{\textwidth}{@{\extracolsep{\fill}}cccccc}
\hline
 $\kappa(\rm GeV^3)$    &0.02&0.04&0.06&0.08 \\
\hline
 $f_{\Sigma_Q^{(\ast)}}(\rm GeV)$    &0.262$\sim$0.288&0.295$\sim$0.299&0.305$\sim$0.336&0.317$\sim$0.361 \\
\hline
 $f_{\Xi_Q^{S(\ast)}}(\rm GeV)$      &0.313$\sim$0.403&0.381$\sim$0.422&0.428$\sim$0.442&0.452$\sim$0.460\\
\hline
 $f_{\Omega_Q^{(\ast)}}(\rm GeV)$    &0.350$\sim$0.507&0.423$\sim$0.542&0.452$\sim$0.559&0.473$\sim$0.571 \\
\hline
\hline
\end{tabular*}
\end{table*}

Generally, the parameters in our model can be determined through the comparison between theoretical predictions and experimental data for $\Lambda_Q$-type or $\Sigma_Q$-type baryons. In Refs.~\cite{Guo:1999ss,Guo:1996jj,Guo:1998at,Guo:2007qu,Guo:1998ef}, some phenomenological predictions for $\Lambda_Q$-type and $\Sigma_Q$-type baryons such as semileptonic, nonleptonic and strong decay widths are given in the BS equation approach. Furthermore, because the heavy quark is not infinite in reality, the $1/m_Q$ corrections to the BS equation for $\Lambda_Q$-type baryons are analyzed in Ref.~\cite{Guo:1999ss} to obtain more accurate predictions. From the analysis of Refs.~\cite{Guo:1999ss,Guo:1996jj,Guo:1998at,Guo:2007qu,Guo:1998ef}, we find the uncertainties of the predictions caused by the parameter $\kappa$ are bigger than those of the diquark mass. However, from the predictions for the residues of $\Lambda_Q$-type and $\Sigma_Q$-type baryons, we find the uncertainties from the mass of diquark ($m_D$ and $m_A$) are larger than those of $\kappa$.

\section{Summary and discussion}
The residues of $\Lambda_Q$-type baryons ($\Lambda_Q$ and $\Xi_Q^A$) and $\Sigma_Q$-type baryons ($\Sigma_Q^{(\ast)}$, $\Xi_Q^{S(\ast)}$ and $\Omega_Q^{(\ast)}$) are interesting quantities both theoretically and experimentally, since they play an important role in calculating the production amplitudes of heavy baryons and can be used to calculate fragmentation functions of heavy baryons. In this work, we constructed the baryonic currents for $\Lambda_Q$-type and $\Sigma_Q$-type baryons in the quark-diquark picture, and then defined the residues of $\Lambda_Q$-type and $\Sigma_Q$-type baryons in this picture. After that, we derived the relations between the residues and corresponding BS wave functions of $\Lambda_Q$-type and $\Sigma_Q$-type baryons.

The BS equations of $\Lambda_Q$ and $\Sigma_Q^{(\ast)}$ were solved numerically based on the interaction kernel including confinement and one gluon exchange terms in the covariant instantaneous approximation. For the doublet states in $\Sigma_Q$-type baryons, for example, $\Sigma_Q$ and $\Sigma_Q^\ast$, we proved that their residues defined in our model are the same under the heavy quark limit, which is consistent with the heavy quark symmetry. Using the numerical solutions for the BS wave functions, we obtained numerical results for the residues, $0.103\,\rm GeV\sim0.224\,\rm GeV$ for $\Lambda_Q$, $0.143\,\rm GeV\sim0.215\,\rm GeV$ for $\Xi_Q^A$, $0.262\,\rm GeV\sim0.361\,\rm GeV$ for $\Sigma_Q^{(\ast)}$, $0.313\,\rm GeV\sim0.460\,\rm GeV$ for $\Xi_Q^{S(\ast)}$ and $0.350\,\rm GeV\sim0.571\,\rm GeV$ for $\Omega_Q^{(\ast)}$. The uncertainties of the residues come from $\kappa$ and the masses of diquarks ($m_D$ and $m_A$). Much more data will be available in the further experimental measurements, which can help us constrain the model parameters more accurately by the comparisons between experimental data and the BS equation predictions.
\section{ACKNOWLEDGEMENT}

This work was supported by the National Natural Science Foundation of China (Projects No.~11775024 and No.~11575023).


\begin{thebibliography}{}
\bibitem{Ebert:1995fp}
  D.~Ebert, T.~Feldmann, C.~Kettner and H.~Reinhardt,
  Z.\ Phys.\ C {\bf 71}, 329 (1996).

\bibitem{Wang:2015kua}
  Z.~G.~Wang,
  Eur.\ Phys.\ J.\ C {\bf 75}, no. 8, 359 (2015).

\bibitem{Groote:1996em}
  S.~Groote, J.~G.~Korner and O.~I.~Yakovlev,
  Phys.\ Rev.\ D {\bf 55}, 3016 (1997).

\bibitem{Guo:1998ef}
  X.~H.~Guo, A.~W.~Thomas and A.~G.~Williams,
  Phys.\ Rev.\ D {\bf 59}, 116007 (1999).

\bibitem{Guo:1996jj}
  X.~H.~Guo and T.~Muta,
  Phys.\ Rev.\ D {\bf 54}, 4629 (1996).

\bibitem{Artru:1989zv}
  X.~Artru and M.~Mekhfi,
  Z.\ Phys.\ C {\bf 45}, 669 (1990).

\bibitem{Meyer:1990fr}
  H.~Meyer and P.~J.~Mulders,
  Nucl.\ Phys.\ A {\bf 528}, 589 (1991).

\bibitem{Tong:1999qs}
  S.~P.~Tong, Y.~B.~Ding, X.~H.~Guo, H.~Y.~Jin, X.~Q.~Li, P.~N.~Shen and R.~Zhang,
  Phys.\ Rev.\ D {\bf 62}, 054024 (2000).

\bibitem{Guo:1992tt}
  X.~H.~Guo and P.~Kroll,
  Z.\ Phys.\ C {\bf 59}, 567 (1993).

\bibitem{Guo:2007qu}
  X.~H.~Guo, K.~W.~Wei and X.~H.~Wu,
  Phys.\ Rev.\ D {\bf 77}, 036003 (2008).

\bibitem{Shuryak:1981fza}
  E.~V.~Shuryak,
  Nucl.\ Phys.\ B {\bf 198}, 83 (1982).

\bibitem{Grozin:1992td}
  A.~G.~Grozin and O.~I.~Yakovlev,
  Phys.\ Lett.\ B {\bf 285}, 254 (1992).

\bibitem{Yakovlev:2000uc}
  O.~I.~Yakovlev, R.~Ruckl and S.~Weinzierl,
  hep-ph/0007344.

\bibitem{Groote:1997ci}
  S.~Groote,
  In *Rostock 1997, Progress in heavy quark physics* 195-198
  [hep-ph/9710365].

\bibitem{GomshiNobary:2009zz}
  M.~A.~Gomshi Nobary, T.~Osati and Z.~Bahadori,
  Nucl.\ Phys.\ A {\bf 821}, 210 (2009).

\bibitem{GomshiNobary:2007ofo}
  M.~A.~Gomshi Nobary, B.~Nikoobakht and J.~Naji,
  Nucl.\ Phys.\ A {\bf 789}, 243 (2007).

\bibitem{Liu:2018tqe}
  L.~L.~Liu, C.~Wang, Y.~Liu and X.~H.~Guo,
  Phys.\ Rev.\ D {\bf 95}, no. 5, 054001 (2017).

\bibitem{Eichten:1978tg}
  E.~Eichten, K.~Gottfried, T.~Kinoshita, K.~D.~Lane and T.~M.~Yan,
  Phys.\ Rev.\ D {\bf 17}, 3090 (1978)
  Erratum: [Phys.\ Rev.\ D {\bf 21}, 313 (1980)].

\bibitem{Wei:2015gsa}
  K.~W.~Wei, B.~Chen and X.~H.~Guo,
  Phys.\ Rev.\ D {\bf 92}, no. 7, 076008 (2015).

\bibitem{Salpeter:1951sz}
  E.~E.~Salpeter and H.~A.~Bethe,
  Phys.\ Rev.\  {\bf 84}, 1232 (1951).

\bibitem{Lurie:1968zz}
  D.~Lurie,
  ``Particles and Fields''.

\bibitem{Jin:1992mw}
  H.~Y.~Jin, C.~S.~Huang and Y.~B.~Dai,
  Z.\ Phys.\ C {\bf 56}, 707 (1992).

\bibitem{Dai:1993np}
  Y.~B.~Dai, C.~S.~Huang and H.~Y.~Jin,
  Phys.\ Lett.\ B {\bf 331}, 174 (1994).

\bibitem{Dai:1993kt}
  Y.~B.~Dai, C.~S.~Huang and H.~Y.~Jin,
  Z.\ Phys.\ C {\bf 60}, 527 (1993).

\bibitem{Georgi:1990cx}
  H.~Georgi,
  Nucl.\ Phys.\ B {\bf 348}, 293 (1991).

\bibitem{Falk:1991nq}
  A.~F.~Falk,
  Nucl.\ Phys.\ B {\bf 378}, 79 (1992).

\bibitem{Guo:1994fn}
  X.~H.~Guo, H.~Y.~Jin and X.~Q.~Li,
  Phys.\ Rev.\ D {\bf 53}, 1153 (1996).

\bibitem{Anselmino:1987vk}
  M.~Anselmino, P.~Kroll and B.~Pire,
  Z.\ Phys.\ C {\bf 36}, 89 (1987).

\bibitem{Mannel:1991ii}
  T.~Mannel, W.~Roberts and Z.~Ryzak,
  Phys.\ Lett.\ B {\bf 271}, 421 (1991).

\bibitem{Roberts:1992xm}
  W.~Roberts,
  Nucl.\ Phys.\ B {\bf 389}, 549 (1993).

\bibitem{Hussain:1993qd}
  F.~Hussain, G.~Thompson and J.~G.~Korner,
  hep-ph/9311309.

\bibitem{Dai:1993kp}
  Y.~B.~Dai, X.~H.~Guo and C.~S.~Huang,
  Nucl.\ Phys.\ B {\bf 412}, 277 (1994).

\bibitem{Guo:2007tn}
  X.~H.~Guo and H.~K.~Wu,
  Phys.\ Lett.\ B {\bf 654}, 97 (2007)

\bibitem{Patrignani:2016xqp}
  C.~Patrignani {\it et al.} [Particle Data Group],
  Chin.\ Phys.\ C {\bf 40}, no. 10, 100001 (2016).

\bibitem{Guo:1999ss}
  X.~H.~Guo, A.~W.~Thomas and A.~G.~Williams,
  Phys.\ Rev.\ D {\bf 61}, 116015 (2000)

\bibitem{Guo:1998at}
  X.~H.~Guo,
  Mod.\ Phys.\ Lett.\ A {\bf 13}, 2265 (1998)
\end{thebibliography}

\end{document}